# Dynamic Load Altering EV Attacks Against Power Grid Frequency Control

Mohammad Ali Sayed, Mohsen Ghafouri, Mourad Debbabi, Chadi Assi
Concordia University, Montreal, Quebec, Canada

*Abstract*— **Driven by the necessity to combat climate change, Electric Vehicles (EV) are being deployed to take advantage of their ability in reducing emissions generated by the transportation sector. This deployment has left the power grid vulnerable to attacks through the EV infrastructure. This paper is written from an attacker's perspective and proposes a dynamic load-altering strategy through manipulating EV charging to destabilize the grid. The attack is formulated based on feedback control theory, i.e., designing attack based on Linear Matrix Inequalities (LMIs). After the stability metric and controller design have been established, we demonstrate our attack method against the Kundur 2-area grid. The attack scenario includes a cap of 200 MW EV load controlled by the attacker. However, the results show that even with this limitation, the attacker would be successful in pushing the grid toward instability and blackout.**

## I. INTRODUCTION

THE power grid is the backbone of modern society, and its security is directly linked to the national security and economic stability of any country. To this end, smart technologies have been introduced to support efficient and reliable grid operation. These technologies, however, pose a security risk, linking the security of the smart grid to the security of the new devices and technologies it utilizes [1].

One such technology is the Electric Vehicle (EV) and its charging infrastructure. The world is witnessing a push for mass deployment of EVs, fueled by the necessity to combat climate change and EVs' potential to reduce the transportations sectors' emissions. In fact, countries have set targets to reach 30% EV sales by 2030 [2], and in 2020 10 million EVs were on the road The exponential growth of EVsis demonstrated by a record 3.2 million new EV sales in 2020 and the expected 6.4 million in 2021 [3]. To support this growth and improve EV charging availability, communication networks and Internet-of-Things (IoT) smart charging has been adopted. This has exposed EV charging to the vulnerabilities of these technologies, making it a security threat to the power grid. Operators have also been hastily deploying charging stations to meet the increasing charging demands which has contributed to the EV ecosystem's lack of proper security. The work presented in [4] demonstrated some of the weaknesses of charging stations and discussed how attackers can gain control of EV charging stations with ease.

In an increasingly interconnected world, cyber-attacks are all but inevitable. The Mirai Botnet [5] is a clear example of the weaknesses inherent to IoT devices that were used by attackers to compromise 600,000 devices to launch denial of service attacks. Recently, the Quebec government announced that their cyber systems are being targeted by well-organized, well-financed groups [6]. Recent studies have considered the disruption of the power grid by compromised high-wattage IoT devices such as water heaters and air conditioners [7]. However, the Black IoT attack [7], and most attacks in the literature, are static and focus on abrupt changes in the load to disrupt the grid. In this paper, we create a dynamic attack that determines the size and the trajectory of the compromised load to achieve grid instability. Furthermore, our attack targets the consumption profile and the actual load instead of targeting the utility or faking sensor measurements like False Data Injection (FDI) attacks. To develop our attack, we rely on cyber-physical attacks and control theory. The contributions of this paper are:

- Modeling the power grid as a feedback control system with the EV load being the feedback input to this system.
- Using Linear Matrix Inequalities (LMIs) and stability metrics to determine the EV attack behavior needed to destabilize a grid.

The rest of the paper is organized as follows. Section II briefly presents the EV ecosystem, its vulnerabilities and related work. Section III presents the methodology and theory of our attack, and the simulation results are presented in Section IV. Finally, Section V offers concluding remarks.

## II. BACKGROUND AND RELATED WORK

In this paper, we present the impact of using EVs as an attack surface to disrupt power grid operation. To that end, this section presents a brief overview of the components of this ecosystem, its vulnerabilities, and a few related studies.

### A. EV Ecosystem Components

*1) EV Charging Stations (EVCSs):* While 11 kW EVCSs are the most common, charging rates range from slow Level 1 chargers with a 1.4kW rate, to 240 kW Level 3 superchargers.

*2)* The EV management system is a cloud-based software that manages the charging sessions and other EVCS functionalities.

*3)* User web/phone applications are used to communicate with the management system and give users control over the EVCs.

*4)* EV protocols are used to achieve bilateral communication between the different components of the EV ecosystem.

*5)* Power Grid: The power grid is the source of electricity for the EVCSs and its security is our largest concern.

### B. EV Ecosystem Vulnerabilities

According to the U.S. Department of Transport [8], the EV industry lacks cybersecurity best practices, and the developed infrastructure does not adhere to a set of standardized/secure communication protocols. It also lacks sufficient protection against cyber intrusions and has no physical security measure.

Multiple studies have demonstrated the vulnerabilities of this ecosystem that allow cyber intrusions and we mention a few:

1) The OCPP protocol used to communicate between the EV management system and EVCSs uses simple HTTP to manage the charging sessions. Also, the optional TLS encryption scheme has mostly been ignored by operators. OCPP's inadequate security leaves it vulnerable to Man-in-the-Middle (MitM) attacks that can highjack charging sessions.

2) The EVCSs host firmware that acts as their management software and is controlled through user applications. As IoT devices, these charging stations are vulnerable to the same attacks as other IoT systems. Multiple works have studied these vulnerabilities and found several security risks [9] [10] in the EV ecosystem. To list a few vulnerabilities that can allow the attackers to manipulate charging sessions, we mention SQL injection, XML injection, Cross-Site Scripting, and missing authentication [9]. Vulnerabilities that have impacts on the user are considered outside of the scope of this paper.

### C. Related Work

The work presented by Soltan et al. [7] demonstrated how a botnet of compromised high-wattage devices can be used to launch large-scale attacks against power grids. The presented attacks are split into attacks against frequency stability, attacks causing tie-line failures, and attacks leading to cascading failure. The authors emphasized that only a small fraction of the available water heaters and air conditioners can be enough to disrupt grid operation and cause blackouts. While their findings are interesting, their attack scenarios were all based on static attacks that rely on a single spike in load. On the contrary, our paper will consider dynamic attack scenarios where the grid is driven into the unstable region by a changing load profile based on attacks designed using feedback control techniques. Also, an EV represents a larger load than traditional high-wattage IoT devices giving attackers a strong attack surface against the grid.

Another form of attack is the switching attack [11]. Switching attacks are still done in an on/off manner to cause another form of instability which is inter-area oscillation.

The work presented in [12], studies the EV attack surface that we consider in this paper. The authors of [12] also represented the power grid in state-space form and considered the EV load to be a feedback controller. However, their choice to utilize the simple full state feedback controllers is simple as compared to our method and may not be optimal due to the unrealistic inputs required by such controllers. Their most important contribution lies in the demonstration of a grid reconnaissance technique based on publicly available data to allow attackers to reconstruct the grid for designing their attacks.

Finally, the authors of [13] provided a comparative table of different stability standards used by different utilities. These stability features and their relation to our attack formulation are discussed below in Sections III and IV.

## III. METHODOLOGY AND ATTACK MODEL

In the following section, we discuss how attackers can build a state-space representation of the power grid based on the grid topology and generators' dynamic behavior. This representation is necessary for the design of smart attack vectors through EVs to destabilize the grid. To achieve this, attackers must have some knowledge about the grid topology.

### A. Power Grid Reconnaissance

Attackers can gain insider information on the grid topology or can even use insiders to compromise a system. The Stuxnet attack against Iranian nuclear facilities in 2010 shows how malware can be introduced into the system by an insider. Malware can also be delivered by compromised software updates like the 2020 SolarWinds hack that infected users such as the U.S. Department of Energy and Department of Homeland Security through a compromised Orion software update. Also, multiple studies have been conducted to estimate power grid topology without the need for such compromises. These methods are designed to recreate the grid topology through:

- Monitoring grid response to load perturbation and performing optimization to determine the parameters that best fit this response [14].
- Machine learning approximation of parameter or line impedances based on PMU measurements [15].

The availability of this data allows the attackers to design the state space representation of the power grid.

### B. State Space and Feedback Control

A well-established concept in control system design is the representation of systems in the form of state space where the variation of states and outputs is governed by (1).

$$\dot{\hat{x}} = A.\hat{x} + B.u$$
$$\hat{y} = C.\hat{x} + D.u \quad (1)$$

Where $\hat{x}$, $\hat{y}$ and $u$ represent the system states, the system output and the system inputs respectively.

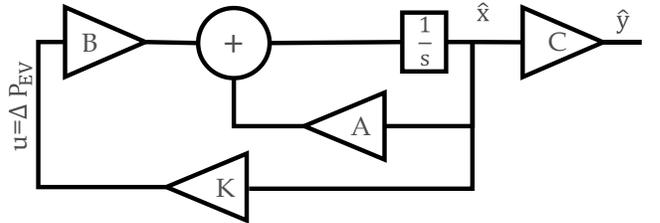

Figure 1. State-space representation of linearized systems.

Feedback controllers as demonstrated by $u = K.\hat{x}$ in Fig. 1, are gain controllers that are designed to determine the required input $u$ for a system based on its current states $\hat{x}$. As such, the EV load $\Delta P_{EV}$ would be determined by the attacker based on $\Delta P_{EV} = u = K.\hat{x}$. The idea behind designing the attack as a feedback is to change the behavioral properties of the grid by changing matrix A which defines the stability conditions of a system based on its eigenvalues. A feedback controller would change the state-space representation into (2) where $A_{cl}=A+B.K$ thus dictating the system behavior by our EV load.

$$\dot{\hat{x}} = A_{cl}.\hat{x} \quad (2)$$

### C. System Modeling

It is understood that a power system's dynamic behavior comes mostly from the generator turbines. Due to their nature as rotating machines, their reactions are not instantaneous and are guided by internal control mechanisms. Fig. 2 represents a simplified model of a generator turbine where the input is considered as the variation in the electric active power load ($\Delta P_{elec}$) at the generator terminal.

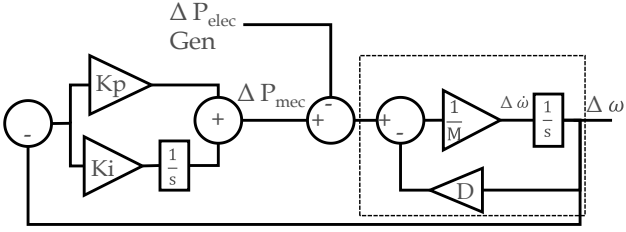

Figure 2. Generator turbine block diagram.

The output is the variation in angular speed $\Delta\omega$. M and D represent the moment of inertia and damping coefficient of the generator respectively. The Kp and Ki are the proportionality and integrator gain of the PI controller. The right area inside the dashed rectangle in Fig. 2 represents the swing equation governing the relation between mechanical power and active power for a turbine. The left half of the block diagram in Fig. 2 is the PI controller responsible for bringing the turbine back to synchronous speed following any disturbance.

Without going into the detailed calculations, (3) represents the state-space representation of a single generator and a connected load. The load is the input of this system and the generator bus angle $\delta$ and speed $\omega$ are the system states.

$$\begin{pmatrix}\dot{\delta}\\ \dot{\omega}\end{pmatrix} = \begin{pmatrix} 0 & 1 \\ \frac{-1}{M}Ki & \frac{-1}{M}(Kp+D)\end{pmatrix} \times \begin{pmatrix}\delta\\ \omega\end{pmatrix} + \begin{pmatrix}0\\ \frac{-1}{M}\times \Delta P_{elec}\end{pmatrix} \quad (3)$$

To expand this model to become a representation of the entire grid, we need to incorporate the power flow equations. To do this we use the linearized DC power flow method that can determine the active power flow while neglecting the reactive power flow and system losses. We consider that the active power flow is sufficient for this study since we are studying load frequency control as a function of active power. The power flowing from bus a to bus b takes the form of (4) and the power injection at bus i is represented by (5):

$$P_{a\text{-}b} = y_{ab} \times (\theta a - \theta b) \quad (4)$$

$$P_i^L = -\sum_{k\in B} Yik\,(\theta i - \theta k) \quad (5)$$

wher $\theta$ is the bus angle, $y_{ab}$ is the admittance of the line, $Yik$ is the admittance matrix of the grid, and B is the set of all buses.

To incorporate the power flow into our state-space, we split buses having generators and loads into generator buses $\in$ B_G and load buses $\in$ B_L. The new power injection at the load and generator buses respectively become (6) and (7).

$$P_i^L = -\sum_{k\in B\_L} Yik\,(\theta i - \theta k) - \sum_{k\in B\_G} Yik\,(\theta i - \delta k) \quad (6)$$

$$P_j^G = \sum_{k\in B\_G} Yjk\,(\delta j - \delta k) + \sum_{k\in B\_L} Yjk\,(\delta j - \theta k) \quad (7)$$

Equation (8) represents the behavior of the load balance at load buses, where $P_{f\_i}$ is the frequency insensitive load and $D_L \times \emptyset$ is the frequency-sensitive load.

$$P^L = P_{f\_i} - D_L \times \emptyset - \Delta P_{EV} \quad (8a)$$
$$\emptyset = \dot{\theta} = 2\pi f \quad (8b)$$

After combining (3), (6), (7), and (8) the extended state-space representing the entire power grid becomes:

$$\begin{pmatrix}\dot{\delta}\\ \dot{\omega}\\ \dot{\theta}\end{pmatrix} = \begin{pmatrix} 0 & I & 0 \\ -M^{-1}(Ki+Y_{GG}) & -M^{-1}(Kp+D) & -M^{-1}Y_{GL} \\ D_L^{-1}Y_{LG} & 0 & D_L^{-1}Y_{LL}\end{pmatrix} \\ \times \begin{pmatrix}\delta\\ \omega\\ \theta\end{pmatrix} + \begin{pmatrix}0\\ 0\\ D_L^{-1}\end{pmatrix} \times \begin{pmatrix}0\\ 0\\ \Delta P_{EV}\end{pmatrix} \quad (9)$$

where L and G are the numbers of load and generator buses, $Y_{GG} \in R^{GxG}$, $Y_{GL} \in R^{GxL}$, $Y_{LL} \in R^{LxL}$ and $Y_{LG} \in R^{LxG}$ are the admittance matrices. $M \in R^{GxG}$, $D \in R^{GxG}$, $D\_L \in R^{LxL}$, $Ki \in R^{GxG}$, $Kp \in R^{GxG}$ are diagonal matrices. $\Delta P_{EV} \in R^{1xL}$ is the input.

### D. Attack Design

After the state-space representation has been established, we consider the design of the feedback controller in this subsection. The design we will present is based on control design through LMIs [16]. LMIs are convex optimization problems with linear matrix constraints. Multiple fields of study utilize LMIs including control system design. In our study, we utilize LMIs to calculate the value of the attack gain K used by the attacker to force the eigenvalues of the system into the desired region. In our dynamic attack, the intent is to relocate the eigenvalues of the system to the region of instability while minimizing the input as will be discussed in the sections below. The regions of instability we selected are based on grid stability standards followed by different utilities such as the New England ISO [13]. These standards specify $\zeta$ and $\omega n$ which are the damping ratio and natural oscillation frequency of a grid. The stable system eigenvalues can be written in terms of $\zeta$ and $\omega n$ as:

$$\lambda_s = a \pm jb \quad (10a)$$
$$a = -\zeta\,\omega n \quad (10b)$$
$$b = \omega n\sqrt{1-\zeta^2} \quad (10c)$$

For the attacker to successfully destabilize the power grid, the eigenvalues of the original system must be shifted into the region of instability which is determined based on $\lambda_{attack} > \lambda_s$. North American utilities, consider a damping ratio of 3-5% or higher as an acceptable stable performance metric for their grids. In this case study, we design our compromised system eigenvalues based on the ISO New England standard that states stability is achieved when $\zeta \geq 3\%$ and $\omega n \geq 0.4\,Hz$ (2.513 rad/s) [13] . Through 10, this is translated into the eigenvalue $\lambda_s = -0.0754 \pm j\,2.512$ beyond which the system is unstable. It is evident from this eigenvalue, that power utilities consider their grid unstable even before the eigenvalue increases beyond. This is to demonstrate the sensitivity and importance of maintaining a stable power grid, due to the technical considerations and the socio-economic damages that occur in case of blackouts.

This is achieved through an LMI technique, called D-Stabilization [16]. After the equations have been written, we solve them using the MATLAB LMI. Fig. 3 is a demonstrative figure with generic poles used to explain the techniques we explain herein.

1) *H(α,β) Stabilization* [16] is a pole placement technique used to design feedback controllers such that the eigenvalues of $A_{cl}$ adhere to $\{a \pm jb \mid -\beta < a < -\alpha\}$ as demonstrated between the two dotted lines in Fig. 3. There can only be a solution to this equation if there exists a symmetric positive definite matrix P satisfying:

$$\begin{cases} (A + BK)P + P(A + BK)^T + 2\alpha P < 0 \\ -(A + BK)P - P(A + BK)^T - 2\beta P < 0 \end{cases} \quad (11)$$

Equation (11) is linearized and rewritten as (12) by assuming $K = WP^{-1}$ and solving for W and P.

$$\begin{cases} AP + PA^T + BW + W^T B^T + 2\alpha P < 0 \\ -AP - PA^T - BW - W^T B^T - 2\beta P < 0 \end{cases} \quad (12)$$

2) *D(q,r) Stabilization* [16] is a pole placement technique used to design feedback controllers such that the eigenvalues of $A_{\_cl}$ adhere to $\{a \pm jb \mid (x + q)^2 + y^2 < r\}$ as highlighted within the dashed circle of Fig. 3. There can only be a solution to this equation if there exists a matrix P satisfying (13).

$$\begin{bmatrix} -rP & qP + (A + BK)P \\ qP + P(A + BK)^T & -rP \end{bmatrix} < 0 \quad (13)$$

Equation (13) is linearized and rewritten as (14) by assuming $K = WP^{-1}$ and solving for W and P.

$$\begin{bmatrix} -rP & qP + AP + BW \\ qP + PA^T + W^T B^T & -rP \end{bmatrix} < 0 \quad (14)$$

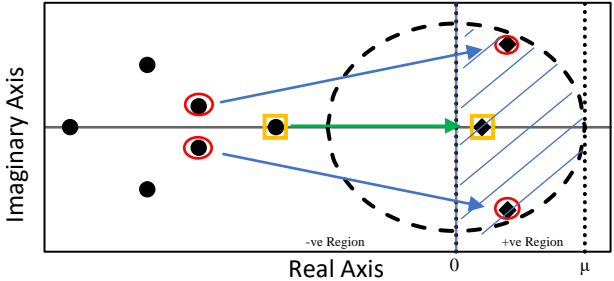

Figure 3. Sample of pole-placement using D-stabilization methods.

These LMI techniques can be used independently or can be combined to achieve a more constrained result in a narrower region based on the attacker's desired behavior. Fig. 3 serves to demonstrate the placement of the poles within the shaded region governed by the rules $H_{(-\mu,0)}$ and $D_{(0,\mu)}$. The eigenvalues represented by the ● represent generic stable poles having a negative real part. These poles are then moved by $\Delta P_{EV}$ to the shaded region based on the gain K. A tight region around the zero axis is chosen for the attack eigenvalues for two reasons. First, having a large eigenvalue would dictate very rapid growth of $\Delta P_{EV}$. The attacker might not control enough EVs to launch such an attack successfully. Secondly, large eigenvalues would cause the system output to increase rapidly, raising alarms for the system operator and exposing the attack.

## IV. CASE STUDY: ATTACK AND DISCUSSION

This case study utilizes the proposed methodology to launch EV attacks against the Kundur 2-area grid demonstrated in Fig. 4 [17]. For the sake of compactness, we will assume the attackers have a fairly accurate representation of the state-space representation of the system. We achieve this by extracting the state-space representation of the system from a MATLAB Simulink model. The Kundur grid has 11 buses, 4 generators, 2 compensators and two loads having a total of 2736 MW.

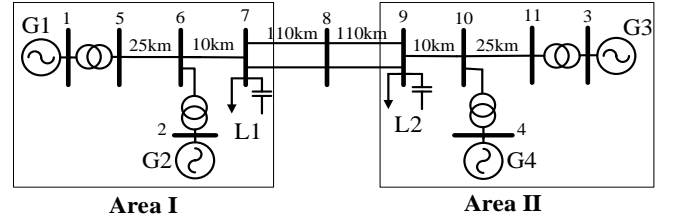

Figure 4. Kundur 2-area power grid.

Based on the state-space model and the calculated attack gain K, the EV load that will be injected into the system is equal to $\Delta P_{EV} = u = K.\hat{x}$. The values used for our attack and the results are presented in the following section.

### A. Attack Implementation and Results

Based on the methodology presented above, we design our attack vector by selecting the region below a=0.5 and centered around zero as the location for the post-attack eigenvalues. The calculated value of K achieved unstable behavior with the most dominant modes: $\lambda_1 = 0.1766$, $\lambda_{2,3} = 0.0095 \pm j\, 0.0687$.

The EV load was modeled using the Simulink Dynamic load with its load values set to an external source that can be both positive and negative to accommodate the EV charging and discharging into the Grid (V2G). Based on feedback control theory, the oscillation of the two EV attack loads is governed by the continuously changing $\hat{x}$ which distinguishes our attack from attacks based on one load spike. As such our EV attack will follow the trajectory determined by $\Delta P_{EV} = u = K.\hat{x}$ as demonstrated in Fig. 5. The unstable system's output will grow exponentially until the grid protection mechanisms trip leading to loss of power and blackouts. In our attack scenario, we capped the compromised EV load at each of the two load busses to $-100\ MW < \Delta P_{EV} < 100\ MW$ as seen in Fig. 5.

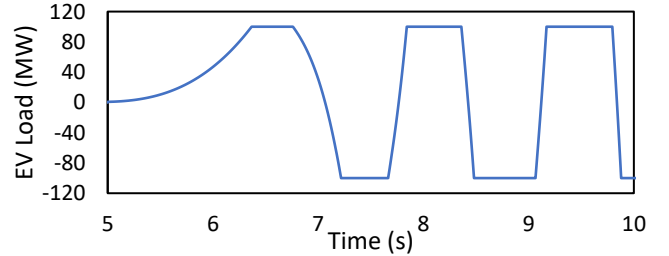

Figure 5. First 5 seconds of the EV attack load at bus 9.

Fig. 6 represents generator 4 behavior after an attack was launched at t=5s. As evident from the frequency graph, the turbine speed deviated significantly from 60 Hz and started oscillating. This unstable behavior was experienced by the 4 generators and is consistent with the achieved eigenvalues.

### B. Result Discussion

Our attack resulted in the destabilization of the system leading to wild oscillation of the generators as demonstrated in Fig. 6. The system goes out of synchronism after t=26s seconds and this would lead the operators to trip the protection system to

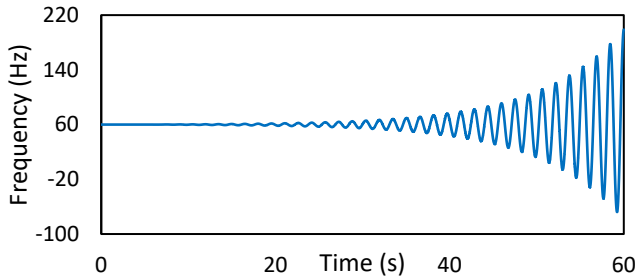
Figure 6. Output frequency of generator 4.

avoid damage to the generators. This happens since the frequency values go above the 5% deviation limit after which generators are instantly tripped [18]. It is worth mentioning, that if we did not cap our load to $\Delta P_{EV} < \pm 100\ MW$ at each load bus, the output oscillations would be larger, and the system experience a blackout earlier as will be demonstrated in the subsection that follows. However, by capping our attack load (Fig. 5), we demonstrate that the attack is still possible with a limited number of EVs under the attacker's control. The total compromised load represents only 7.3% of the system load but was enough to destabilize the system. The total 200MW load is equal to 18,000 EVs charging at 11 kW or 900 EVs charging at 240 kW superchargers, or any combination in between.

### C. Different Attack Conditions

Further experiments were conducted to demonstrate the impact of the attackers' ability to compromise different numbers of EVs. The different $\Delta P_{EV}$ loads and the times at which the system reaches the 2.5% and 5% frequency limits are presented in Table I. Table I also contains the result of a simulation where the attackers designed the trajectory of the EV attack load based on inaccurate information of the grid topology. We simulate this by adding 10% random error to the values of the state-space matrices.

TABLE I. TABLE TYPE STYLES

| Test Case | Time Needed to Reach Frequency Deviation Thresholds | |
|---|---|---|
| | 2.5% Threshold | 5% Threshold |
| $\Delta P_{EV} \leq \pm 50\ MW$ | 25.1 sec | 31.3 sec |
| $\Delta P_{EV} \leq \pm 100\ MW$ | 19.5 sec | 25.5 sec |
| $\Delta P_{EV} \leq \pm 200\ MW$ | 15.5 sec | 20.1 sec |
| $\Delta P_{EV} \leq \pm 100\ MW$ with 10% error in parameters | 20.8 sec | 26.9 sec |

The values in Table I demonstrate that compromising larger EV loads, causes faster system instability and blackouts. It is also important to mention when the attack load trajectory is designed with imperfect grid knowledge, the instability is still achieved under the same load conditions albeit slightly delayed. This can be attributed to the gain K value being optimized based on the wrong set of parameters making it less optimal when used against the accurate system.

## V. CONCLUSION

In this work, we presented how EVs can be used to recreate the behavior of feedback loop controllers to destabilize a power system. This study is conducted from the perspective of the attacker that built a state-space representation of the power grid including the dynamic behavior of the generator turbines. The attacker then uses that model to design the feedback gain based on Linear Matrix Inequalities such as the D-Stabilization methods to cause unstable grid operation to cause blackouts. This was achieved by compromising an EV load equal to 7.3% of the total system load.

REFERENCES

[1] S.-Che, H. Adam, L. Ching, "Cyber security of a power grid: state-of-the-art," Int J Electr Power Energy Syst, vol 99, pp 45–56, 2018.
[2] IEA (2020), "Global EV Outlook 2020," IEA, Paris https://www.iea.org/reports/global-ev-outlook-2020.
[3] "Global Plug-in Vehicle Sales Reached over 3,2 Million in 2020," http://www.ev-volumes.com/.
[4] H. ElHussini, C. Assi, B. Moussa, R. Atallah, and A. Ghrayeb, "A Tale of Two Entities Contextualizing Electric Vehicle Charging Stations Cyber-Fingerprint on the Power Grid", ACM Transactions on Internet of Things, vol. 2, no. 2, pp 1-21, 2021.
[5] M. Antonakakis, T. April, M. Bailey, M. Bernhard, E. Bursztein, J. Cochran, Z. Durumeric, J. A., Invernizzi, L., M. Kallitsis, et al, "Understanding the Mirai botnet," USENIX Security Sympsion'17, pp 1093-1110, 2017.
[6] "States are funding cyberattacks on Quebec systems, says minister," https://montreal.ctvnews.ca/states-are-funding-cyberattacks-on-quebec-systems-says-minister-1.5607331
[7] S. Soltan, P. Mittal and H. V. Poor, "BlackIoT: IoT botnet of high wattage devices can disrupt the power grid,' 27th USENIX Security Symposium (USENIX Security 18), pp. 15-32, 2018.
[8] K. Harnett, B. Harris, D. Chin, G. Watson, "DOE/DHS/DOT Volpe Technical Meeting on Electric Vehicle and Charging station Cybersecurity Report," Technical Report. John A. Volpe National Transportation Systems Center (US), 2018.
[9] T. Nasr, S Torabi, E. Bou-Harb, C. Fachkha and C. Assi, "Power Jacking Your Station: In-Depth Security Analysis of Electric Vehicle Charging Station Management Systems" Computers & Security, 2021.
[10] P. Ferrara, A.K. Mandal, A. Cortesi, et al, "Static analysis for discovering IoT vulnerabilities". International Journal on Software Tools for Technology Transfer vol 23, pp 71-88, 2021.
[11] M.E. Kabir, M. Ghafouri, B. Moussa, & C. Assi(2021). A Two-Stage Protection Method for Detection and Mitigation of "Coordinated EVSE Switching Attacks," IEEE Transactions on Smart Grid, vol 12 no 5, pp 4377–4388, 2021.
[12] S. Acharya, Y. Dvorkin and R. Karri, "Public Plug-in Electric Vehicles + Grid Data: Is a New Cyberattack Vector Viable?", IEEE Transactions on Smart Grid, vol. 11, no. 6, pp. 5099-5113, 2020.
[13] M. A. Tabrizi, N. Prakash, M. Sahni, H. Khalilinia, P. Saraf, and S. Kolluri, "Power system damping analysis on large power system networks: An energy case study," in 2017 IEEE Power & Energy Society General Meeting. IEEE, pp. 1–5, 2017.
[14] S. Taheri, V. Kekatos and G. Cavraro, "An MILP Approach for Distribution Grid Topology Identification using Inverter Probing," 2019 IEEE Milan PowerTech, Milan, Italy, pp. 1-6, 2019.
[15] K. Moffat, M. Bariya and A. Von Meier, "Unsupervised Impedance and Topology Estimation of Distribution Networks-Limitations and Tools," IEEE Transactions on Smart Grid, vol. 11, no. 1, pp. 846-856, 2020.
[16] G. Duan and H. Yu, "LMIs in Control Systems: Analysis, Design and Applications," CRC Press, 2013.
[17] "Two Area System," https://electricgrids.engr.tamu.edu/electric-grid-test-cases/two-area-system/
[18] C. Luo, H.G. Far, H. Banakar, P.K Keung,& B.T. Ooi, "Estimation of Wind Penetration as Limited by Frequency Deviation". IEEE Transactions on Energy Conversion, vol. 22, no. 3, pp 783-791, 2007.